\documentclass[aps,prl, twocolumn, superscriptaddress, groupedaddress]{revtex4}

\usepackage{graphicx}  
\usepackage{bm} 
\usepackage{amssymb}  
\usepackage{mhchem}
\usepackage{siunitx}
\usepackage{color}

\begin{document}
\widetext

\title{Observation of oscillatory relaxation in the Sn-terminated surface of epitaxial rock-salt SnSe $\{111\}$ topological crystalline insulator}


\author{Wencan Jin}
\altaffiliation{These authors contribute equally to this work.}
\affiliation{Columbia University, New York, New York 10027, USA}

\author{Suresh Vishwanath} 
\altaffiliation{These authors contribute equally to this work.} 
\affiliation{Cornell University, Ithaca, New York 14853, USA}

\author{Jianpeng Liu} 
\affiliation{Kavli Institute for Theoretical Physics, University of California, Santa Barbara, California 93106, USA }

\author{Lingyuan Kong} 
\affiliation{Beijing National Laboratory for Condensed Matter Physics, and Institute of Physics, Chinese Academy of Sciences, Beijing 100190, China}

\author{Rui Lou} 
\affiliation{Department of Physics, Renmin University of China, Beijing 100872, China}

\author{Zhongwei Dai} 
\affiliation{University of New Hampshire, Durham, NH 03824, USA}

\author{Jerzy T. Sadowski} 
\affiliation{Center for Functional Nanomaterials, Brookhaven National Laboratory, Upton, New York 11973, USA}

\author{Xinyu Liu} 
\affiliation{University of Notre Dame, Notre Dame, Indiana 46556, USA}

\author{Huai-Hsun Lien} 
\affiliation{Cornell University, Ithaca, New York 14853, USA}

\author{Alexander Chaney} 
\affiliation{Cornell University, Ithaca, New York 14853, USA}

\author{Yimo Han}
\affiliation{Cornell University, Ithaca, New York 14853, USA}

\author{Micheal Cao}
\affiliation{Cornell University, Ithaca, New York 14853, USA}

\author{Junzhang Ma} 
\affiliation{Beijing National Laboratory for Condensed Matter Physics, and Institute of Physics, Chinese Academy of Sciences, Beijing 100190, China}

\author{Tian Qian} 
\affiliation{Beijing National Laboratory for Condensed Matter Physics, and Institute of Physics, Chinese Academy of Sciences, Beijing 100190, China}

\author{Jerry I. Dadap} 
\affiliation{Columbia University, New York, New York 10027, USA}

\author{Shancai Wang} 
\affiliation{Department of Physics, Renmin University of China, Beijing 100872, China}

\author{Malgorzata Dobrowolska}
\affiliation{University of Notre Dame, Notre Dame, Indiana 46556, USA}

\author{Jacek Furdyna} 
\affiliation{University of Notre Dame, Notre Dame, Indiana 46556, USA}

\author{David A. Muller}
\affiliation{Cornell University, Ithaca, New York 14853, USA}

\author{Karsten Pohl} 
\affiliation{University of New Hampshire, Durham, NH 03824, USA}

\author{Hong Ding} 
\affiliation{Beijing National Laboratory for Condensed Matter Physics, and Institute of Physics, Chinese Academy of Sciences, Beijing 100190, China}

\author{Huili Grace Xing}
\email{grace.xing@cornell.edu} 
\affiliation{Cornell University, Ithaca, New York 14853, USA}
\affiliation{University of Notre Dame, Notre Dame, Indiana 46556, USA} 

\author{Richard M. Osgood, Jr} 
\email{osgood@columbia.edu}
\affiliation{Columbia University, New York, New York 10027, USA}  

\begin{abstract}  
Topological crystalline insulators have been recently predicted and observed in rock-salt structure SnSe $\{111\}$ thin films. Previous studies have suggested that the Se-terminated surface of this thin film with hydrogen passivation, has a reduced surface energy and is thus a preferred configuration. In this paper, synchrotron-based angle-resolved photoemission spectroscopy, along with density functional theory calculations, are used to demonstrate conclusively that a rock-salt SnSe $\{111\}$ thin film epitaxially-grown on \ce{Bi2Se3} has a stable Sn-terminated surface. These observations are supported by low energy electron diffraction (LEED) intensity-voltage measurements and dynamical LEED calculations, which further show that the Sn-terminated SnSe $\{111\}$ thin film has undergone a surface structural relaxation of the interlayer spacing between the Sn and Se atomic planes. In sharp contrast to the Se-terminated counterpart, the observed Dirac surface state in the Sn-terminated SnSe $\{111\}$ thin film is shown to yield a high Fermi velocity, $0.50\times10^6$m/s, which suggests a  potential mechanism of engineering the Dirac surface state of topological materials by tuning the surface configuration. 
\end{abstract}

\maketitle

\section{Introduction}
\begin{figure*}
\includegraphics[scale=1.5]{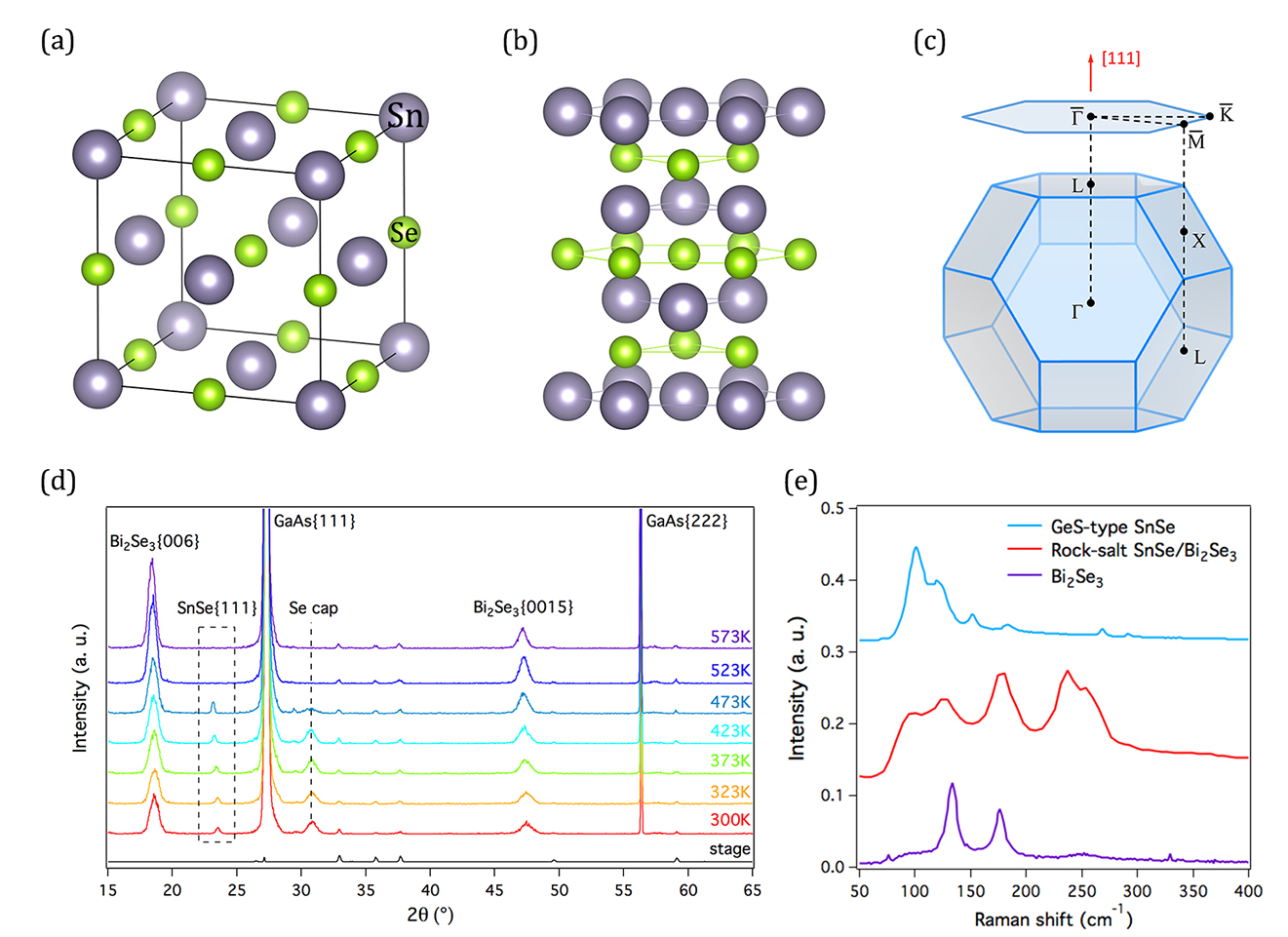}
\caption{\label{fig:fig1} (a) Schematic of the lattice structure of rock-salt SnSe. (b) Layered rock-salt SnSe depicted along the $\{111\}$ direction. (c) The Brillouin zone of bulk rock-salt SnSe and its projected $\{111\}$ surface Brillouin zone. (d) Temperature dependent XRD of SnSe/\ce{Bi2Se3}/GaAs acquired using Cu $K_{\alpha}$ line. Black dashed box and line denote the diffraction peaks for SnSe $\{111\}$ and Se cap, respectively. (e) Raman spectra of GeS-type SnSe (blue), rock-salt SnSe (red), and \ce{Bi2Se3} (purple). The excitation wavelength is 488nm.}
\end{figure*}

Topological phases of matter have attracted much interest for condensed matter physics community. The initial focus and discovery in this area was topological insulators (TIs) \cite{hasan2010colloquium,moore2010birth,qi2011topological}, in which time reversal symmetry gives rise to topologically protected, conducting surface states in insulating bulk crystals. Beyond TIs, the search for new types of topological materials has recently also been extended to other cases, in which the role of topological protection arises from other symmetries including particle-hole symmetry \cite{schnyder2008classification}, magnetic translational symmetry \cite{mong2010antiferromagnetic}, and crystalline symmetry \cite{fu2011topological}. In the last class of topological materials, topologically nontrivial properties are protected by crystalline symmetry, and are thus called topological crystalline insulators (TCIs) \cite{fu2011topological}. The TCI phase has been theoretically predicted and experimentally verified in narrow-gap IV-VI semiconductors: SnTe \cite{tanaka2012experimental,tanaka2013two,littlewood2010band, zhang2016arpes}, \ce{Pb_{1-x}Sn_{x}Se} \cite{polley2014observation} and \ce{Pb_{1-x}Sn_{x}Te} (x$\ge$0.25) \cite{xu2012observation,yan2014experimental,tanaka2013tunability,safaei2013topological,gyenis2013quasiparticle,dziawa2012topological} with a rock-salt structure. In these materials, bulk-band gaps are located at the four equivalent L points of the face-centered-cubic (FCC) Brillouin zone (see Fig.~\ref{fig:fig1}(c)); in addition, the energy-level ordering of the conduction and valence bands at the L points is inverted \cite{hsieh2012topological}. A remarkable signature of the TCI phase is the presence of surface states with an even number of gapless Dirac cones on the $\{001\}$, $\{111\}$ surface Brillouin zone (SBZ), these states are protected by the mirror symmetry with respect to the $\{110\}$ plane \cite{hsieh2012topological,sun2013rocksalt,liu2013two}.\\

For topological materials, in general, the surface configuration is crucial for their Dirac surface states. In particular, tin monochalcogenide TCI is an important platform for investigating the correlation between Dirac surface states and its surface configuration. Since the surface states of TCIs are crystal-symmtry protected, they depend sensitively on the surface orientation. In particular, the $\{001\}$ surface states possess hybridized double-Dirac-cones in close vicinity to the $\bar{X}$ point of the $\{001\}$ SBZ, while the $\{111\}$ surface states possess four Dirac cones centered at the $\bar{\Gamma}$̅ and $\bar{M}$ points of the $\{111\}$ SBZ \cite{tanaka2013two}. In previous experiments, the $\{001\}$ surface states have been more intensively investigated because the $\{001\}$ surface is a natural cleavage plane of rock-salt IV-VI semiconductors. In contrast, the $\{111\}$ surface is a polar surface, which is typically difficult to prepare because of its unstable structure arising from the divergence of the electrostatic energy along polar direction, i.e., the well know \textit{polar catastrophe} \cite{tasker1979stability, noguera2000polar}. This phenomenon may be ameliorated by structural reconfiguration of the surface, which strongly depends on the type of surface termination. Until now, a limited surface termination has been employed in $\{111\}$ surface of rock salt TCI. For example, recent studies of \ce{Pb_{1-x}Sn_{x}Te} have indicated Te-termination of its $\{111\}$ surface \cite{yan2014experimental}. Very recently, TCI phase in metastable SnSe $\{111\}$ thin film with rock salt structure, has been observed and was attributed to be Se-terminated and hydrogen passivated using first-principles calculation \cite{wang2015molecular}. Thus previous experiments have shown the stablity of anion-terminated TCIs, either with Te or Se. The cation-terminated surface, i.e., the truncated-bulk Sn-terminated surface, however, was predicted to be unstable, and, as a result, it undergoes surface reconstruction to diminish its surface energy \cite{wang2014structural}. Despite its great importance to the full understanding of this family of TCIs and its role in the development of potential high-quality TCI devices, a stable cation-terminated TCI has not yet been demonstrated.\\

In the present work, we present clear evidence of a stable, truncated-bulk Sn-terminated SnSe $\{111\}$ TCI, using a comprehensive battery of experimental and theoretical tests. We also present the first direct evidence of oscillatory structural relaxation of the top atomic layers of this TCI. The surface structure of these thin films is characterized using selected-area low-energy electron diffraction ($\mu$-LEED). In addition, LEED intensity-voltage (\textit{I-V}) measurements, combined with dynamical LEED calculations, show that our SnSe $\{111\}$ thin films have a Sn-terminated surface without surface reconstruction. Instead structural relaxation of the interlayer spacing between the Sn and Se atomic planes occurs. Our angle-resolved photoemission spectroscopy measurement demonstrate a robust surface state at the $\bar{\Gamma}$̅ point with a Dirac point located at $\sim$0.4eV below \textit{$E_{F}$}. In addition, these measurements reveal a much higher Fermi velocity (\textit{$v_F$}) in the Sn-terminated SnSe than for its Se-terminated counterpart. The measured bands of Sn-terminated SnSe $\{111\}$ are accurately reproduced by our first-principles calculations. Despite the fact that most topological material design focuses on composition \cite{polley2016observation, lou2015sudden}, this work paves the possibility of obtaining a tunable Dirac point and Fermi velocity in TCI by modifying the surface termination; it also shows clearly one potential approach to manipulation of topologically nontrivial surface states by tuning the surface structure through choice of growth conditions or decapping conditions.\\

\section{Results and discussion}
\begin{figure*}
\includegraphics[scale=0.6]{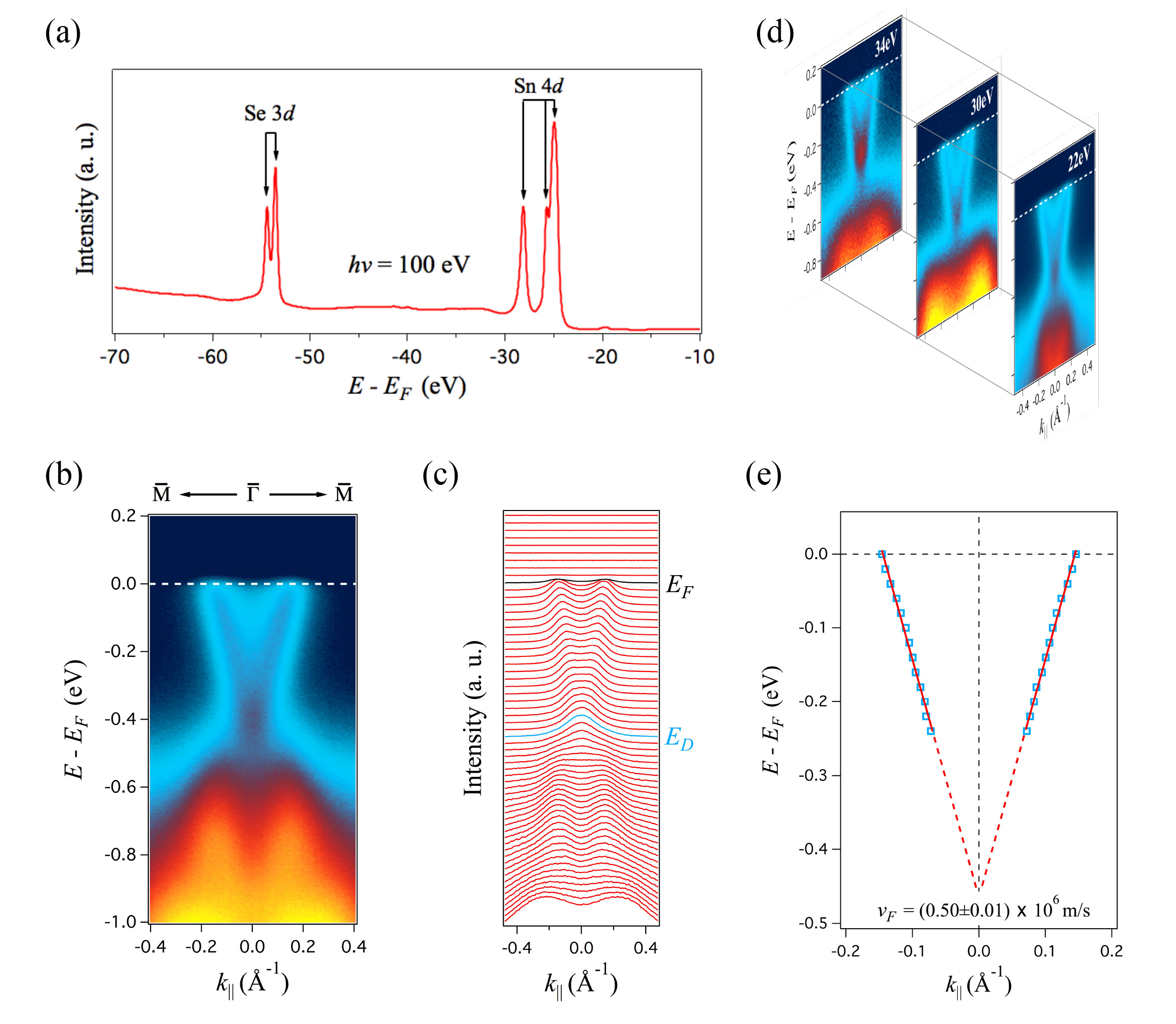}
\caption{\label{fig:fig2} (a) Core level photoemission spectrum acquired using 100eV incident photon energy. (b) ARPES bandmap ($h\nu$=25eV) along the $\bar{M}$-$\bar{\Gamma}$-$\bar{M}$ high symmetry direction. (c) MDCs plot of the band dispersion shown in (b). Energy positions of the Fermi level and Dirac point are denoted as \textit{$E_F$} and \textit{$E_D$}, respectively. (d) Excerpts of the photon-energy-dependent ($h\nu$=22, 30 and 34eV) ARPES bandmaps. (e) MDCs peak positions (blue dots) and linear fitting (red dashed line). }
\end{figure*}

\begin{figure*}
\includegraphics[scale=0.55]{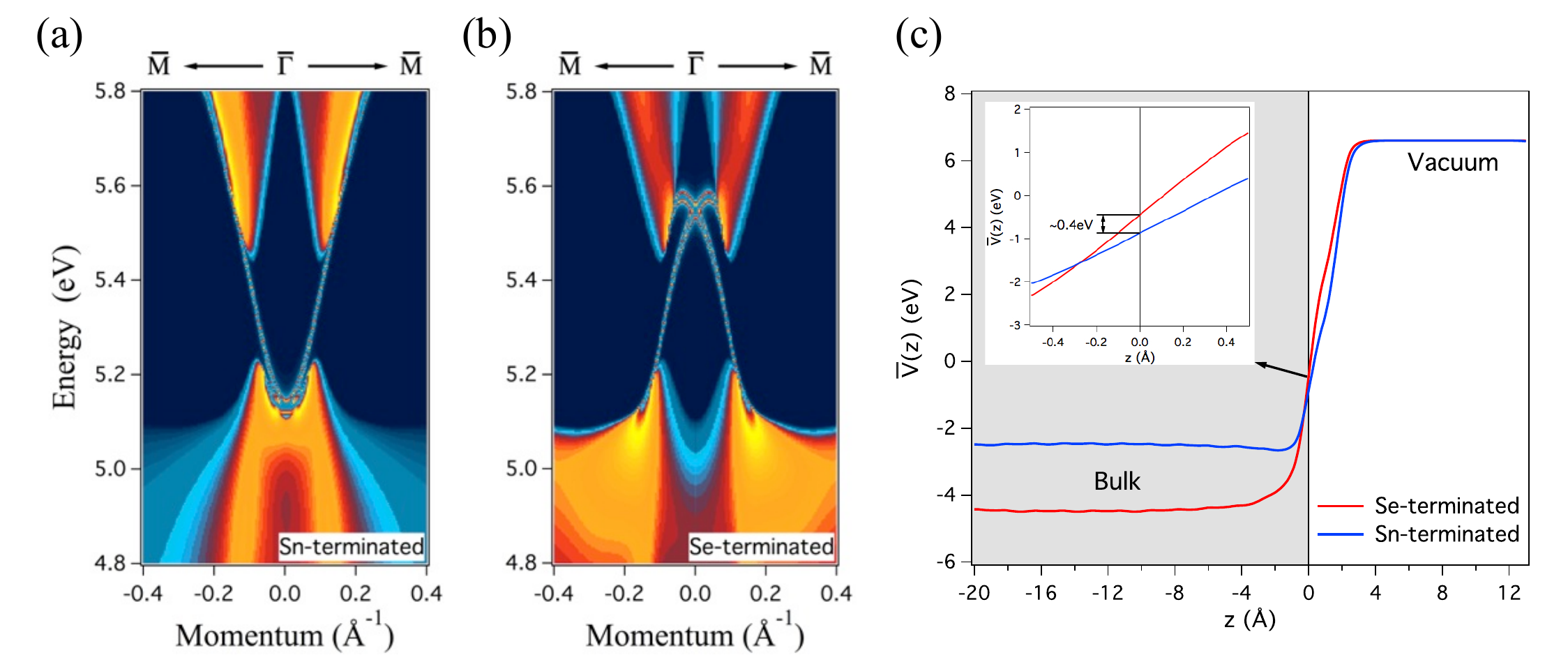}
\caption{\label{fig:fig3} First-principles calculations of the band structure for (a) Sn-terminated and (b) Se-terminated SnSe $\{111\}$ thin film. (c) Average electrostatic energy of Sn-terminated (blue) and Se-terminated (red) SnSe thin film as a function of $z$. Inset shows a magnification of the curves at the surface.}
\end{figure*}

The Sn-Se system contains very rich phases. The energetically stable phase of SnSe has an orthorhombic GeS structure, which is a topologically trivial phase. Molecular beam epitaxy (MBE) is used to grow SnSe thin film in rock-salt structure (see Fig.~\ref{fig:fig1}(a)). This film has a thickness of 26 monolayers and is grown on a crystalline \ce{Bi2Se3} thin film on a GaAs $\{111\}$ substrate. The lattice constant of \ce{Bi2Se3} is closely matched to the in-plane lattice constant of rock-salt SnSe and, as a result, the epitaxial SnSe $\{111\}$ (see Fig.~\ref{fig:fig1}(b)) is constrained to a rock-salt structure. The growth conditions are detailed in the \textbf{Methods} section. The growth was carefully monitored and characterized \textit{in situ} using reflection high-energy electron diffraction (RHEED) (see Supplementary Section I). The SnSe thin film is protected from ambient atomospheric exposure following growth with a Se cap in the growth chamber.  Prior to our microscopy or spectroscopy measurements on each sample, the Se cap was removed by heating at \SI{200}{\degree}C for 30min in ultrahigh vacuum.\\ 

The crystalline structure of the epitaxially grown thin film was first investigated using temperature-dependent X-ray diffraction (XRD) from room temperature (RT) to \SI{300}{\degree}C (see Supplementary Section II for a discussion on off-axis XRD). The results, shown in Fig.~\ref{fig:fig1}(d), exhibit consistently well-defined peaks for GaAs $\{111\}$, GaAs $\{222\}$, \ce{Bi2Se3} $\{006\}$, and \ce{Bi2Se3} $\{0015\}$. In addition, well-defined peaks for SnSe $\{111\}$ and Se cap are seen. Note that in both on-axis and off-axis XRD measurements, no peak for orthorhombic SnSe is observed. Also, the Se cap peaks started to break down at \SI{200}{\degree}C, while the SnSe $\{111\}$ peaks slightly shifted due to thermal expansion and remained intact until \SI{200}{\degree}C. This result suggests that heating the sample at \SI{200}{\degree}C can successfully remove the Se cap without damaging the SnSe thin film. In addition, ancillary Raman spectra, which are shown in Fig.~\ref{fig:fig1}(e), support our supposition that this SnSe layer has a rock salt structure, which is clearly distinct from the structure of GeS-type SnSe. Finally, note that the well-resolved SnSe and \ce{Bi2Se3} peaks seen on the as-grown sample indicate that no significant intermixing during growth \cite{vishwanath2016controllable}.\\

As mentioned above, there have been no reports of a direct measurement of the surface structure of rock-salt SnSe prior to the present work. Also, since rock-salt SnSe is a metastable phase, we found that transmission electron microscopy, which use high-energy electron beam bombardment of samples, triggered a transformation in SnSe and at the SnSe/Se cap interface (see Supplementary Section III). Similar electron-beam induced transformation in Sn-Se system was also reported recently in exfoliated \ce{SnSe2} \cite{sutter2016electron}. In order to establish this surface structure, measurements were carried out on our SnSe thin film using $\mu$-LEED and LEED \textit{I-V}. The resulting SnSe $\mu$-LEED pattern shows one set of sharp hexagonal diffraction spots along with a faint ring-like background. This result is consistent with our thin films having one dominant in-plane crystal orientation along with a small percentage of randomly misaligned small domains. This observation allows us to rule out the ($2\times1$) or $\sqrt{3}\times\sqrt{3}$ R\SI{30}{\degree} reconstruction \cite{wang2014structural}.\\ 

Additional measurements were made of the sample surface (see Supplementary Sections IV and V). First of all, a measurement was made of any variation in the domains along the surface. The probing electron beam (\SI{5}{\micro\meter} in diameter) was translated across the sample for several hundreds of micrometers. No obvious change in the $\mu$-LEED pattern was observed, indicating long-range domain uniformity of the thin film. Secondly, surface roughness was quantitatively characterized using variations in the local surface normal (see Supplementary Section III), and the result indicated that epitaxial SnSe and \ce{Bi2Se3} layers were atomically flat. Moreover, in order to enable acquisition of the $\mu$-LEED pattern of GaAs $\{111\}$ substrate, each of the epitaxial thin layers were successively removed \textit{in situ} by heating the sample slowly to \SI{400}{\degree}C. The GaAs $\{111\}$ substrate was found to display a well-defined hexagonal diffraction pattern, which was used to provide a known reciprocal lattice template by assuming the known lattice constant of GaAs as \SI{5.65}{\angstrom} \cite{kaminska1989structural}. The in-plane lattice constant of SnSe thin film was then determined to be 4.28$\pm$\SI{0.03}{\angstrom} by using the reciprocal vector ratio extracted from our $\mu$-LEED patterns. This value for SnSe is $\sim1\%$ larger than the published measured value of \SI{4.23}{\angstrom} \cite{mariano1967polymorphism}, yet it is in good agreement with the structural optimized value in density functional theory (DFT) calculations \cite{sun2013rocksalt,wang2015molecular}.\\

The electronic structure of the SnSe $\{111\}$ sample was directly measured using synchrotron-based high-resolution ARPES system. Figure~\ref{fig:fig2}(a) shows the integrated core-level photoemission spectrum of the sample. Well-defined Se 3$d$ and Sn 4$d$ spin-orbit doublets are immediately apparent in this figure. The peak at 28eV is assigned to a replica Sn 4$d$ state originating from the chemical shift of the Sn surface. (see Supplementary Section VI for a photoionization cross section analysis.) Note that since 26 layers of SnSe is much thicker than the penetration depth of the low energy photon, it is unlikely that the surface states of the \ce{Bi2Se3} buffer layer will be observed. Figure~\ref{fig:fig2}(b) shows the ARPES bandmap along the $\bar{M}$-$\bar{\Gamma}$-$\bar{M}$ high-symmetry direction acquired using a 25eV incident photon energy. Figure~\ref{fig:fig2}(c) shows the momentum distribution curves (MDCs) plot of the bandmap of Fig.~\ref{fig:fig2}(b). In sharp contrast to the electronic structure of Se-terminated SnSe with hydrogen-passivation, where Dirac point at $\bar{\Gamma}$̅ point is located at approximately 0.1 eV below Fermi level (\textit{$E_{F}$}), we observed Dirac-like linear dispersive bands crossing at $\sim$0.4eV below \textit{$E_F$}. In addition, to verify the $k_z$ ($\{111\}$ direction in BZ) dispersion of the band features, photon-energy-dependent ARPES measurements were carried out (see Supplementary Section VIII). Figure~\ref{fig:fig2}(d) shows the excerpts from the photon-energy-dependence (h$\nu$=22, 30 and 34eV) ARPES bandmap. The Dirac-like band feature does not show any noticeable change with varying incident photon energy over a wide energy range, confirming that it is a surface state. However, the bands with binding energy higher than 0.5eV do show strongly photon-energy-dependent evolution, indicating their bulk origin. As shown in Fig.~\ref{fig:fig2}(e), linear fitting to the MDC peaks yields a Fermi wave vector of $k_F$=0.14$\pm$\SI{0.01}{\per\angstrom} and a high Fermi velocity of $v_{F}$=(0.50$\pm$0.01)$\times10^6$m/s. The Fermi velocity for Sn-terminated SnSe sample is three times larger than for its Se-terminated counterpart.\\

To aid in interpreting the electronic structure and topological character, first-principles calculations were carried out (see \textbf{Methods} section for calculation details). In contrast to the method reported in Ref.\cite{wang2015molecular}, in which  dangling bonds were eliminated using hydrogen-passivation of the Se-termination, a truncated-bulk surface was used in our calculations. In fact in our calculation, the Sn-terminated surface (see Fig.~\ref{fig:fig3}(a)) and Se-terminated surface (see Fig.~\ref{fig:fig3}(b)) are found to yield strikingly different surface states. In the Sn-terminated case, the Dirac point is close to the bulk valence band, while in the Se-terminated case, the Dirac point is close to the bulk conduction band. This result implies that our SnSe $\{111\}$ thin film has a Sn-terminated surface. In fact, a linear fitting to the calculated surface state of the Sn-terminated SnSe yields a Fermi velocity of 0.55$\times10^6$m/s, which is in good agreement with our measured value. The physical origin of the Dirac point shift is attributed to the different electrostatic energy of these two surfaces. In particular, this $z$-dependent average electrostatic energy is obtained using the formula:\\

$\bar{V}(z)=\frac{1}{cA}\int_{z+c/2}^{z-c/2}dz\int\int_{A}dxdyV(x,y,z)$

where $V(x,y,z)$ is the microscopic electrostatic energy, $c$ is the lattice constant in the $z$ direction ($z$ is the normal of SnSe $\{111\}$ surface), and $A$ is the area of the unit cell in x-y plane. The blue (red) curve in Fig.~\ref{fig:fig3}(c) is the average electrostatic energy $\bar{V}$(z) for the Sn-terminated (Se-terminated) slab. Even though $\bar{V}$(z) of the Se-terminated slab is lower than that of the Sn-terminated slab in the bulk region, the electrostatic energy increases rapidly at the surface. As shown in the inset of Fig.~\ref{fig:fig3}(c), at the surface indicated by the vertical gray line, $\bar{V}$(z) of the Sn-terminated slab is lower than that of the Se-terminated slab by $\sim$0.4eV, which is consistent with our ARPES data and first-principles calculations.\\

\begin{figure}
\includegraphics[scale=0.4]{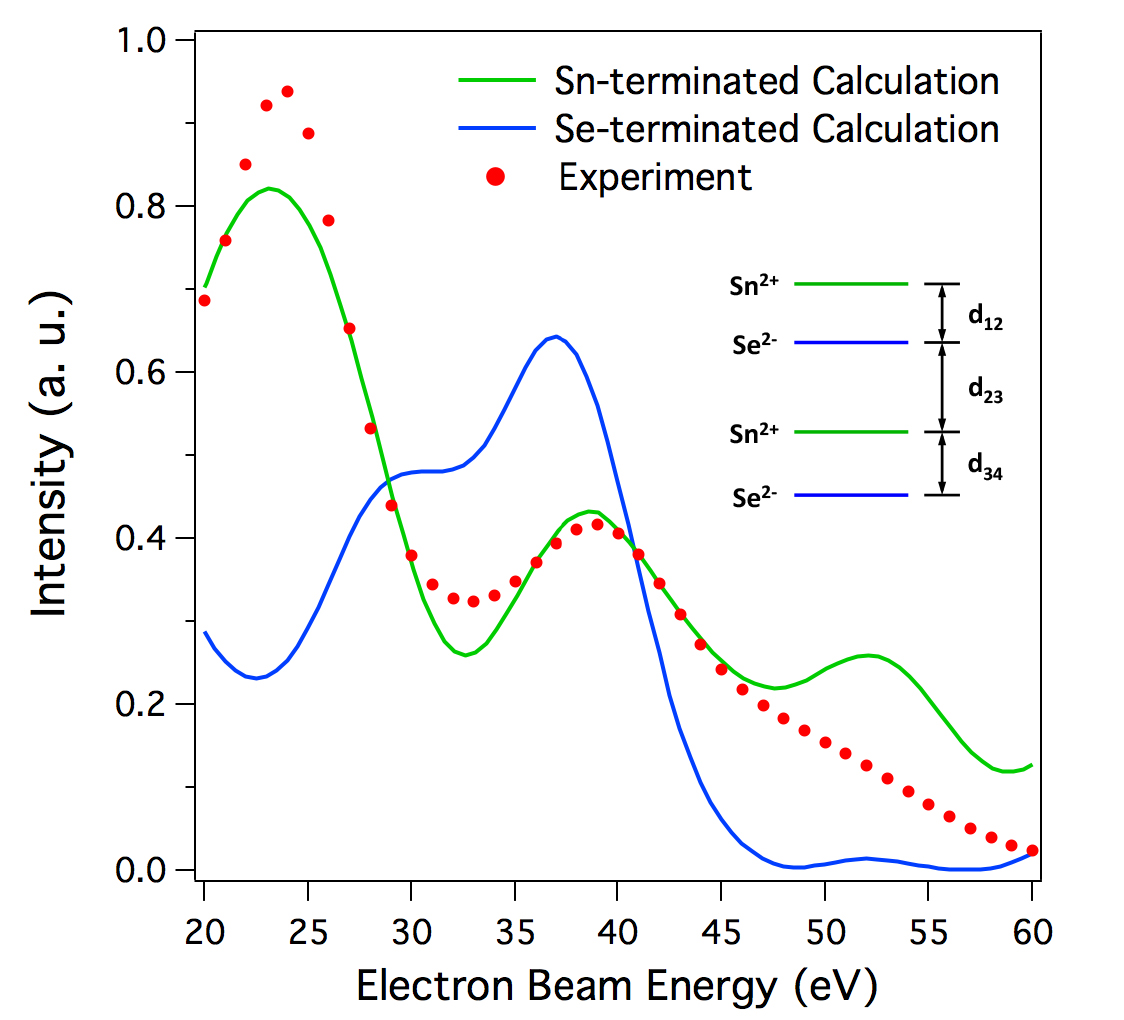}
\caption{\label{fig:fig4} Calculated LEED \textit{I-V} curves for the (00) diffraction beam for an optimized Sn-terminated surface (green solid curve) and a Se-terminated surface (blue solid curve) and measured electron reflectivity curve (red dots).}
\end{figure}

The surface termination can also be independently confirmed using $\mu$-LEED \textit{I-V} analysis. Specifically, LEED \textit{I-V} measurements were carried out to extract the energy dependence of the electron reflectivity of the (00) diffraction beam \cite{dai2017surface}. Also, \textit{I-V} curves were calculated using dynamical multiple scattering codes \cite{adams2002simple} for different trial structures \cite{sun2010spatially}. As a result, the local surface structure can be determined by comparing the experimental reflectivity curve with calculated \textit{I-V} curves. As shown in Fig.~\ref{fig:fig4}, the calculated \textit{I-V} curve of an optimized Sn-terminated surface accurately reproduces the major features of our measured \textit{I-V} curve, while the calculated \textit{I-V} curve of a Se-terminated surface is strikingly different from the experimental data. This result strongly supports a SnSe thin-film surface that is Sn-terminated.\\ 

\begin{table}[h]
\caption{Calculated optimum top few layers spacings $d_{ij}$ between the \textit{ith} and \textit{jth} atomic plane (the inset of Fig. 4) for a SnSe thin film with Sn-terminated surface and the relaxation with respect to the bulk layer spacing $d_0$.}
\centering
\begin{tabular} { c c c } 
 \hline
Parameters & Optimized values (\SI{}{\angstrom}) & $\triangle d_{ij}/d_{0}$ ($\%$) \\ [0.75ex] 
 \hline
 $d_{12}$ & 1.64 & -6.2 \\
 \hline
 $d_{23}$ & 1.89 & 8.1 \\ 
 \hline
 $d_{34}$ & 1.68 & -4.1 \\ 
 \hline
\end{tabular}
\label{table:1}
\end{table}

An additional important question is the stability of the polar surface of SnSe. As shown in Fig.~\ref{fig:fig1}(b), a SnSe $\{111\}$ thin film has a stacking sequence of \ce{Sn^{2+}} and \ce{Se^{2-}} atomic planes, which give rise to a surface dipole moment and surface charge \cite{tasker1979stability}. Such a stacking sequence can be compensated through the formation of a suitable surface reconstruction \cite{wang2014structural}; however, such a compensation mechanism has been ruled out in our case through our LEED measurement. Another possible charge-compensation mechanism is a spatial variation of the slabs along the dipole direction \cite{noguera2000polar}. Therefore, in our structural-optimization procedure, the top four layers of the SnSe $\{111\}$ thin film with Sn-termination are permitted to adjust their interlayer spacing around their initial value of \SI{1.75}{\angstrom}, thus enabling compensation of charge. As a result of this calculation, best-fit parameters are obtained and summarized in Table I, revealing an oscillatory, contraction-expansion-contraction pattern for the structural relaxation in the top few layers of SnSe. This is the first direct evidence of the oscillatory structural relaxation predicted by DFT calculation \cite{wang2014structural}. Note that a good agreement between calculated and experimental \textit{I-V} curves is achieved in the energy range of 20-60eV (electron penetration depth in this range is estimated to be $<$\SI{8}{\angstrom}), suggesting the structural relaxation is occuring in the top few layers. However, due to the limited accessible energy range (20-100eV), a meaningful final R-factor was not quantified here.\\
 
\section{Conclusion}
In conclusion, we have performed comprehensive experimental and theoretical investigation of the surface and electronic structure of an epitaxial rock-salt SnSe $\{111\}$ topological crystalline insulator. This investigation, allows us to demonstrate that our SnSe $\{111\}$ thin film has a pristine Sn-terminated surface, which is stabilized via an oscillatory variation of the spacings between the top few layers. In our ARPES experiments, robust surface states with ultrahigh Fermi velocity are observed at the SBZ center. Such distinct properties may lead to potential applications in electronic and spintronic devices, and open up a possible route to the manipulation of surface states via tuning of the surface termination. 

\section{Methods}
\textbf{Molecular beam epitaxy growth.} MBE growth was carried out with \textit{n}-doped GaAs $\{111\}$ B substrates using a Riber 32 MBE at the University of Notre Dame. Prior to growth, the substrate was annealed gradually to \SI{700}{\degree}C to remove the surface oxide \textit{in situ}. Subsequently, the substrate was exposed to a Se flux of $1.8\times10^{-6}$ Torr at \SI{700}{\degree}C for 15 mins. The substrate was then cooled to \SI{340}{\degree}C and allowed to stabilize at this latter temperature for 40 mins. \ce{Bi2Se3} was then grown at the substrate temperature of \SI{340}{\degree}C for 15 mins under simultaneous incident elemental Bi and Se fluxes of $3.5\times10^{-8}$ Torr and $1.8\times10^{-6}$ Torr, respectively. Based on the known flux, the expected thickness of the \ce{Bi2Se3} layer was 12 monolayers ($\sim$12 nm). Subsequent to this step, the substrate temperature was then lowered to \SI{200}{\degree}C and stabilized for 30 mins. SnSe growth was done at a substrate temperature of \SI{200}{\degree}C for 17 mins under simultaneous incidence of an elemental Sn and an Se flux of $2.1\times10^{-8}$ Torr and $2.4\times10^{-8}$ Torr, respectively. The thickness of SnSe is $\sim$26 monolayers. Se layers were used to cap the as-grown SnSe thin film \textit{in situ} so as to protect the surface from ambient exposure during transport of the sample \cite{vishwanath2015comprehensive,park2016scanning,vishwanath2016controllable}.\\

\textbf{X-ray diffraction} Temperature dependent XRD was done using a Rigaku SmartLab X-ray diffractometer. The Se-capped sample was used for the measurement. The sample stage was made of AlN and the measurement was done in an inert atmosphere of nitrogen.  The temperature was raised at a rate of \SI{5}{\degree}C/min and held at the measurement temperature for 10 mins to stabilize the sample prior to measurement. High resolution of the measurement was achieved by using Ge (022)$\times$4 monochromator on the source end of the X-ray.\\

\textbf{Angle-resolved photoemission spectroscopy.} ARPES measurements were performed at the Dreamline beamline of the Shanghai Synchrotron Radiation Facility (SSRF) with a Scienta D80 analyzer. The samples were decapped in a preparation chamber at \SI{200}{\degree}C and then measured at 40 K in a vacuum with a pressure $<5\times10^{-11}$ Torr. The ARPES data were collected within 12 hours after decapping, during which no signature of surface degradation was observed. The energy and angular resolutions were set to 15 meV and \SI{0.2}{\degree}, respectively.\\

\textbf{Low energy electron microscopy.} $\mu$-LEED measurements were performed at the Center for Functional Nanomaterials, Brookhaven National Laboratory using an ELMITEC AC-LEEM system. In this system, the sample was annealed \textit{in situ} from \SI{200}{\degree}C (for SnSe) to \SI{300}{\degree}C (for \ce{Bi2Se3}), and to \SI{500}{\degree}C (for GaAs), and acquired the $\mu$-LEED in real-time. The spatial resolution is $<$3 nm in LEEM mode. The electron-beam spot size in the $\mu$-LEED mode was \SI{5}{\micro\meter} in diameter.\\ 

\textbf{First-principles electronic structure calculation.} DFT \cite{hohenberg1964inhomogeneous,kohn1965self} calculations of the bulk SnSe electronic structure were performed using a VASP package \cite{kresse1996efficient,kresse1996efficiency}. The generalized gradient approximation (GGA) \cite{perdew1996generalized} was adopted to describe the exchange-correlation potential. Hybrid functional (HSE) was also examined and found yield comparable result as with GGA (see Supplementary Section IX). The in-plane hexagonal lattice parameter was set as $a$=\SI{4.24}{\angstrom}, and the height of the repeating unit of Sn-Se bilayer was set as \SI{3.46}{\angstrom}. The energy cutoff was set to 400 eV. The Brillouin zone was sampled by a $8\times8\times3$ $k$-point mesh. A 48-band tight-binding model was then constructed in Wannier function basis using a Wannier90 package \cite{marzari1997maximally,souza2001maximally,mostofi2008wannier90}, where the Wannier functions were generated by projecting the Bloch functions obtained from DFT calculations above onto the spinor \textit{s} and \textit{p} orbitals located at all the Sn and Se sites. Finally, the surface states were calculated in a semi-infinite geometry using the iterative surface Green's function method as reported in Ref. \cite{sancho1985highly}.\\ 

\textbf{Dynamical LEED Calculation.} The codes from Adams \textit{et al}. \cite{adams2002simple}, which were developed from the programs of Pendry \cite{pendry1977low} and Van Hove and Tong \cite{van2012surface}, were used in the dynamical LEED calculations. The in-plane lattice constant was set to be \SI{4.28}{\angstrom}, a value determined using our $\mu$-LEED pattern. The Debye temperature for SnSe was set as 210K. The inner potential of SnSe is set as 10.1eV. 12 (L=11) phase shifts are used in the calculation. 

\section{Acknowledgements}
\begin{acknowledgements} 
First, the authors acknowledge very helpful discussions with Liang Fu and David Vanderbilt. In addition, we acknowledge Andrew Kummel and Jun Hong Park for development of decapping process. the LEEM/LEED research was carried out in part at the Center for Functional Nanomaterials, Brookhaven National Laboratory, was supported by the U.S. Department of Energy, Office of Basic Energy Sciences, under Contract No. DE-SC0012704. The MBE growth, Raman and XRD characterizations were supported by the NSF EFRI-2DARE project No. 1433490 and NSF Grant DMR 1400432. This work also made use of the Cornell Center for Materials Research Shared Facilities, which are supported through the NSF MRSEC program (DMR-1120296). The DFT calculations were carried out on RUPC. The work of R.M.O., J.D., and W.J. was financially supported by the U.S. Department of Energy under Contract No. DE-FG 02-04-ER-46157. In addition, J.L. was supported by DMR-1408838 and DMR-1506119. Z.W.D., and K.P. were supported by NSF DMR 1006863. R.L. and S.C.W. were supported by the National Natural Science Foundation of China (No. 11274381). L.Y.K., J.Z.M., T.Q., and H.D. were supported by the Ministry of Science and Technology of China (No. 2015CB921300, No. 2013CB921700), the National Natural Science Foundation of China (No. 11474340, No. 11234014), and the Chinese Academy of Sciences (No. XDB07000000).
\end{acknowledgements} 

\bibliography{SnSe.bib}

\newpage
\pagenumbering{gobble}
\clearpage

\widetext
\begin{center}
\textbf{\Large Supplementary materials}\\
\end{center}

\noindent \textbf{I. Reflection high energy electron diffraction (RHEED)}\\

\begin{figure}[!h]
\includegraphics[scale=0.75]{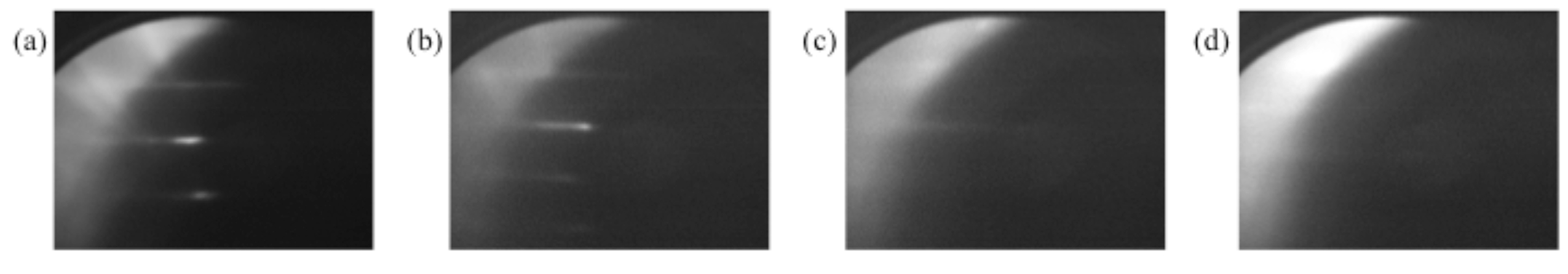}
\renewcommand{\thefigure}{S\arabic{figure}}
\caption{\label{fig:fig.S1}RHEED patterns of (a) GaAs $\{111\}$ before growth and after annealing in the presence of Se flux, (b) after \ce{Bi2Se3} growth, (c) 120 sec SnSe growth ($\sim$3 monolayers) and (d) 1040 sec SnSe growth ($\sim$26 monolayers).}
\end{figure}

Figure~\ref{fig:fig.S1}  shows the streaky RHEED patterns indicating a smooth film growth. The sharp RHEED pattern in Fig.~\ref{fig:fig.S1}(a) demonstrates that the GaAs $\{111\}$ substrate is clean after removing surface oxide and annealing in Se flux. Also, the clear RHEED pattern of \ce{Bi2Se3} epitaxial layer shown in Fig.~\ref{fig:fig.S1}(b) suggests high crystalline quality of the thin film. However, as shown in Fig.~\ref{fig:fig.S1}(c) $\&$ (d), with increasing thickness of SnSe, the RHEED gets weaker. This observation reinforces the $\mu$-LEED observation showing greater randomness in grain orientation as compared to the \ce{Bi2Se3} and single-crystal GaAs $\{111\}$ substrate.

\noindent \textbf{II. Off-axis X-ray diffraction (XRD)}\\

\begin{figure}[!h]
\includegraphics[scale=0.5]{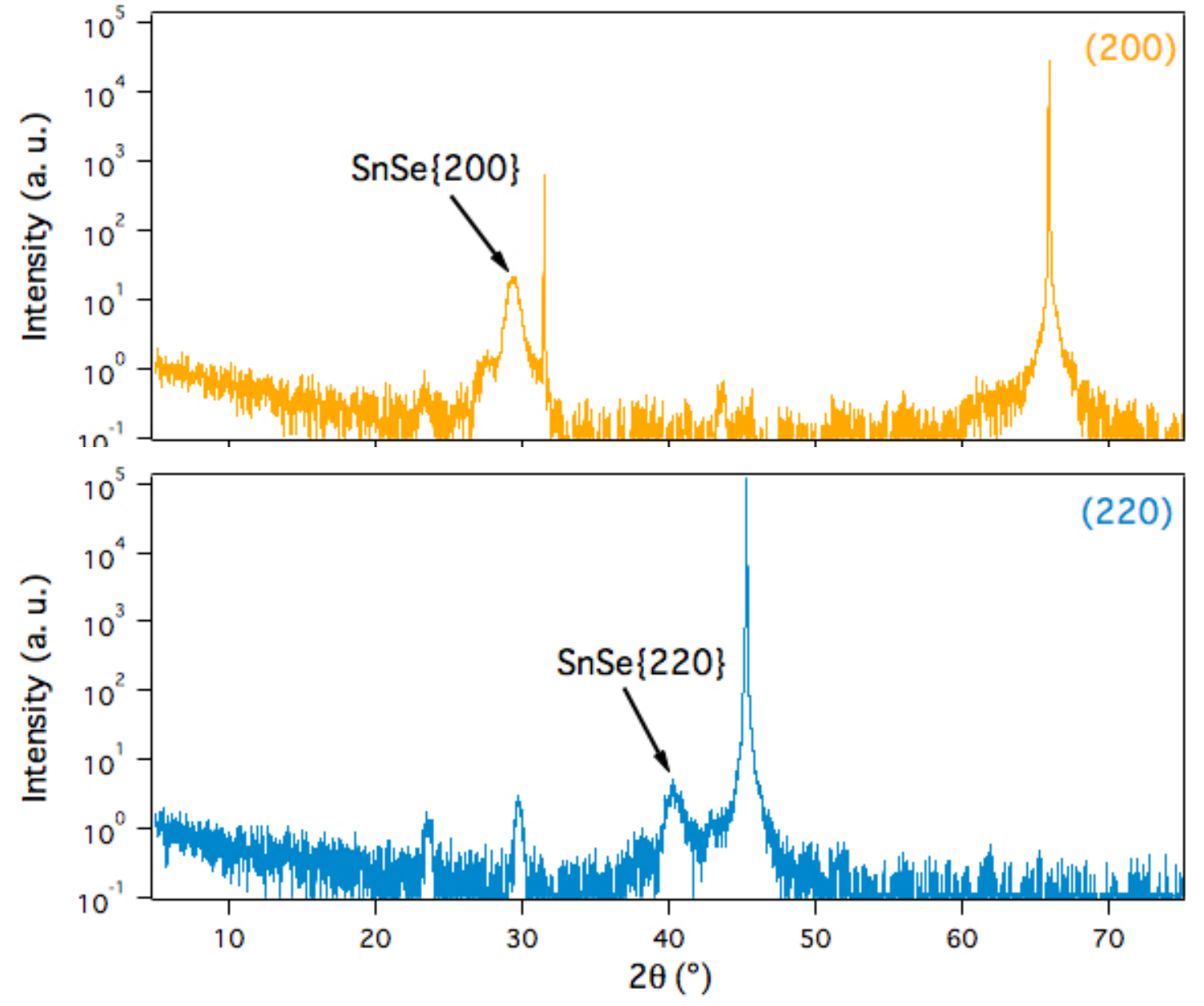}
\renewcommand{\thefigure}{S\arabic{figure}}
\caption{\label{fig:fig.S2} Off-axis XRD scan along (200) (yellow), and (220) (blue), respectively.}
\end{figure}

To verify the crystalline structure of the SnSe $\{111\}$ thin film, we further carried out off-axis XRD measurement. Figure~\ref{fig:fig.S2} shows the off-axis scan along (200) and (220), respectively. As summarized in Table~\ref{table:Table S1}, the peak for $\{200\}$ and $\{220\}$ are assigned to 2$\theta$=\SI{27.20}{\degree} and \SI{38.85}{\degree}, respectively. Our off-axis XRD suggests that our SnSe thin film is slightly distorted.

\begin{table}[!h]
\renewcommand{\thetable}{S\arabic{table}}
\caption{\label{table:Table S1} Off-axis XRD results}
\centering
\begin{tabular} { c  c  c  c } 
 \hline
Peak   & Expected 2$\theta$ (\SI{}{\degree})   &   Measured 2$\theta$ (\SI{}{\degree})  &   Measured d (\SI{}{\angstrom}) \\ [0.75ex] 
 \hline
 $\{200\}$ & 27.20 & 29.37 & 3.04\\ 
 \hline
 $\{220\}$ & 38.85 & 40.35 & 2.24 \\ 
 \hline
\end{tabular}
\end{table} 

\begin{figure}[!h]
\includegraphics[scale=0.35]{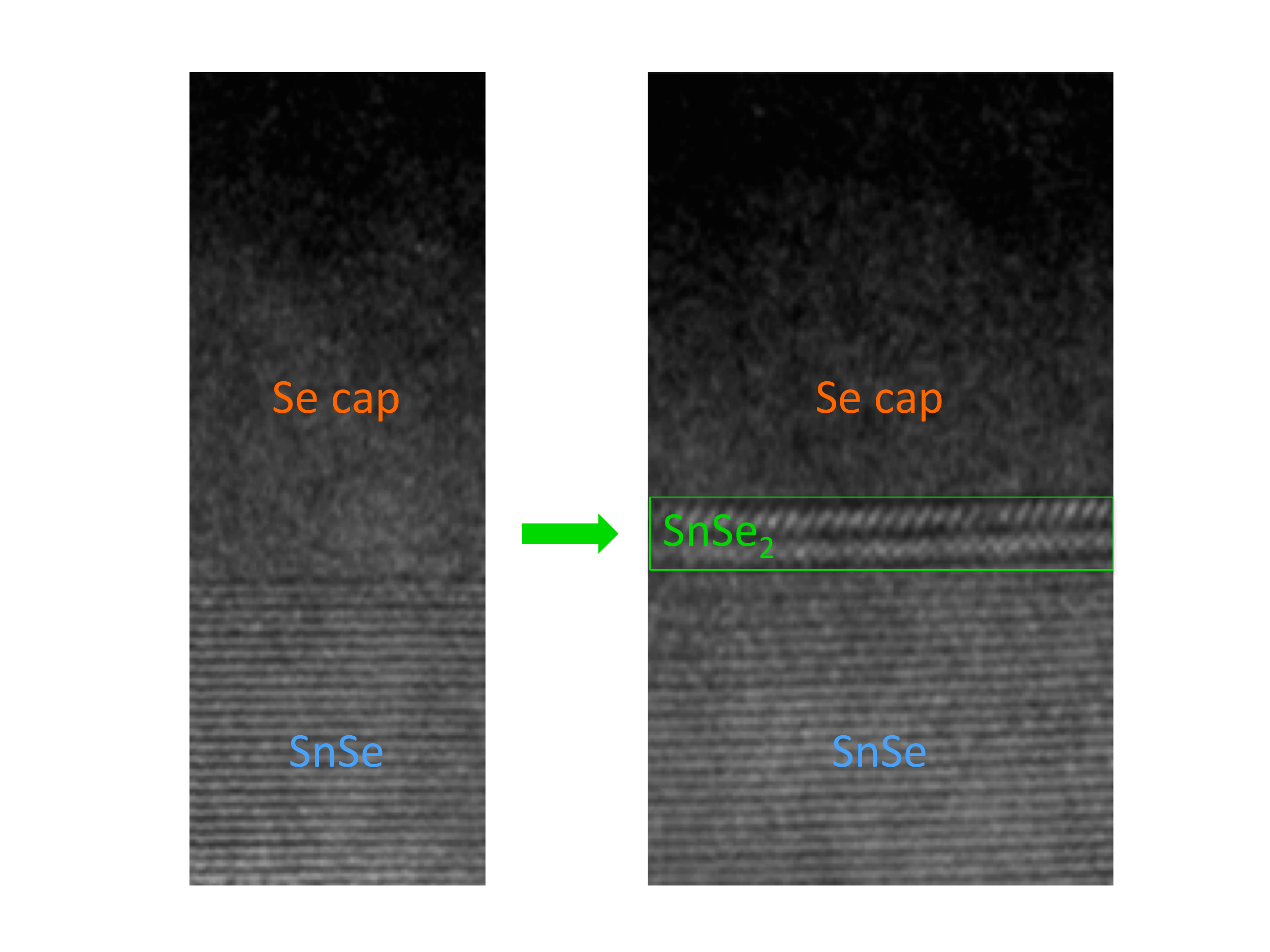}
\renewcommand{\thefigure}{S\arabic{figure}}
\caption{\label{fig:fig.S3} Atomic resolution ADF-STEM images at t=0 (left) and t=10s (right) after bombardment with a high-energy electron beam (100kV). A transformation from SnSe $\{111\}$ (blue) to 2H-\ce{SnSe2} (green) is shown at the interface.}
\end{figure}

\noindent \textbf {III. Transmission electron microscopy (TEM)}\\

{As rock-salt SnSe is a \textit{metastable} phase, transmission electron microscopy (TEM), which uses high-energy electron beam bombardment of samples is found to trigger transformations in Sn-Se system. The TEM image shows GeS-type SnSe in-place of the rock-salt SnSe. Also, e-beam induced transformation from SnSe $\{111\}$ to 2H-\ce{SnSe2} was found at the SnSe/Se cap interface (see Fig.~\ref{fig:fig.S3}). Here, we attack this issue by decapping the sample in ultrahigh vacuum (UHV), and carrying out surface characterization with \textit{low-energy} electron beam as the probe. As shown in Section IV, our results demonstrate a successful method for studying \textit{metastable} materials.\\

\noindent  \textbf{IV. Low energy electron microscopy and diffraction (LEEM and LEED)}\\

Figure~\ref{fig:fig.S4}(a) shows the schematic of our sample configuration (SnSe/\ce{Bi2Se3}/GaAs) with Se cap as a protective layer. Figure~\ref{fig:fig.S4}(b) shows LEEM images of our as-load sample with an amorphous surface of the Se cap. Prior to the measurements, the Se cap is removed by annealing the sample in UHV at \SI{200}{\degree} for 30min. As shown in Fig.~\ref{fig:fig.S4}(c), the SnSe thin film is flat and uniform with sparse residual Se nuclei.\\ 

\begin{figure}[h!]
\includegraphics[scale=0.5]{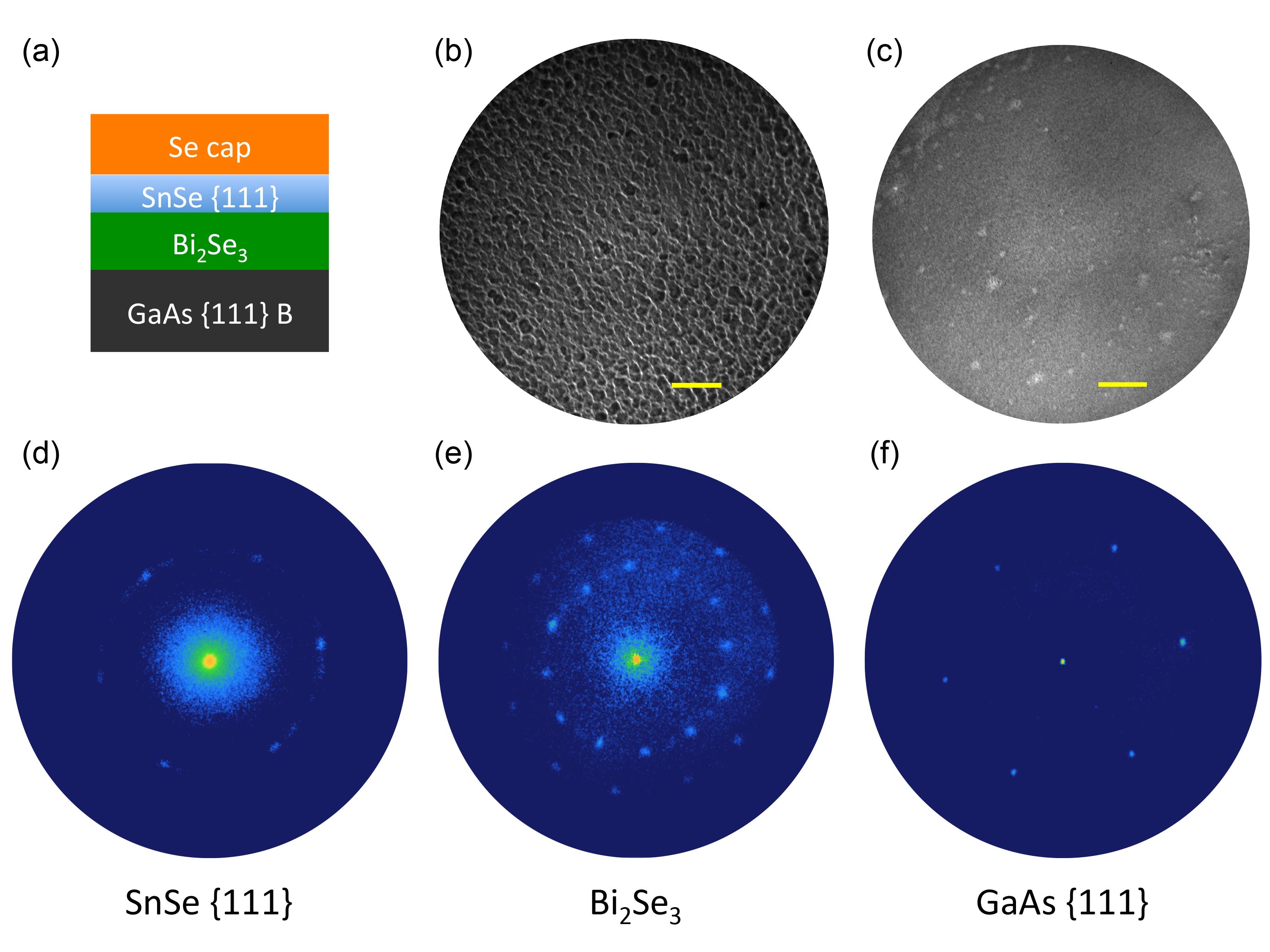}
\renewcommand{\thefigure}{S\arabic{figure}}
\caption{\label{fig:fig.S4}(a) Schematic of the sample configuration. (b) LEEM image of the amorphous surface of Se cap. (b) LEEM image of the SnSe thin film after decapping in UHV. The scale bar is \SI{5}{\micro\meter}. $\mu$-LEED patterns of (c) SnSe $\{111\}$, (d) \ce{Bi2Se3}, and (e) GaAs $\{111\}$.
}
\end{figure} 

\begin{figure}[h!]
\includegraphics[scale=0.5]{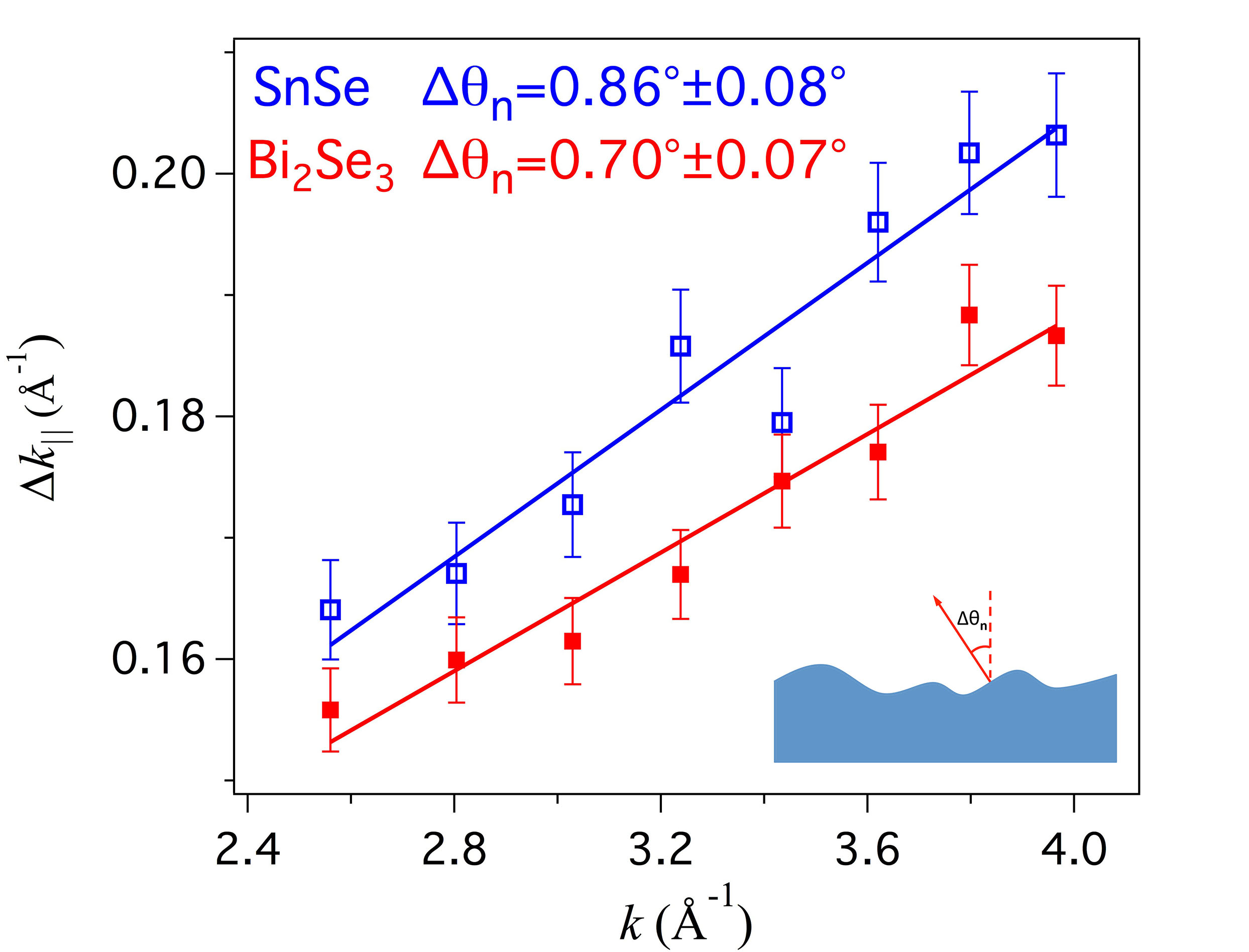}
\renewcommand{\thefigure}{S\arabic{figure}}
\caption{\label{fig:fig.S5}Linewidth of the central diffraction beam of SnSe (blue) and \ce{Bi2Se3} (red) thin film as a function of k. The inset shows the schematic of the local surface normal.}
\end{figure}

To fully investigate the orientation of the thin films and substrates, $\mu$-LEED measurements were carried out. After removing the Se cap, $\mu$-LEED patterns were acquired at room temperature. Subsequently, the sample was gradually heated to \SI{300}{\degree} to remove the SnSe layers, thus giving the $\mu$-LEED pattern of the bare \ce{Bi2Se3} surface layer. After acquiring the $\mu$-LEED patterns of \ce{Bi2Se3} at \SI{300}{\degree}, the sample was heated to \SI{500}{\degree}C to remove the \ce{Bi2Se3} layer and permit measurement of the GaAs substrate $\mu$-LEED pattern. Figure~\ref{fig:fig.S4}(d)-(f) shows the $\mu$-LEED patterns from the SnSe $\{111\}$, \ce{Bi2Se3}, and GaAs $\{111\}$ layers, respectively. These patterns show that the SnSe $\{111\}$ layer has one preferred orientation with a faint ring-like misalignment. The \ce{Bi2Se3} layer has two dominant orientations, rotated by \SI{30}{\degree} relative to each other. GaAs $\{111\}$ surface has a well-defined hexagon $\mu$-LEED pattern with sharp diffraction spots. The reciprocal space are calibrated in real space by assuming that the GaAs $\{111\}$ substrate has a lattice constant of a=\SI{5.65}{\angstrom}. This then sets the in-plane lattice constant of SnSe $\{111\}$ layer to be 4.28$\pm$\SI{0.04}{\angstrom}, i.e. a $\sim$1$\%$ expansion in comparison to the bulk value.\\ 

\noindent \textbf{V. Surface roughness characterization}\\

The surface corrugation can be quantitatively characterized using a model, for which $\Delta\theta=\Delta k_{||}/2k$, where $\Delta\theta$ is the standard deviation of the local surface normal (see the inset of Fig.~\ref{fig:fig.S5}), $k_{||}$ is the linewidth of the central diffraction spots, and $k=\sqrt{2mE_{kin}}$ is the momentum of incident electron. In our experiments, the incident electron energy was tuned from 20-100eV and acquired the corresponding $\mu$-LEED patterns. As shown in Fig.~\ref{fig:fig.S5}, the linewidth of the central diffraction spots increase linearly with $k$. The magnitude of the surface roughness of SnSe $\{111\}$ and \ce{Bi2Se3} thin film is very small ($<$\SI{1}{\degree}), indicating an atomically flat surface.\\

\noindent \textbf{VI. Photoionization cross section}\\

The photon energy we used to acquire ARPES data is between 20 and 70 eV, which is very surface sensitive (penetration depth $\sim$2-\SI{10}{\angstrom}). As 26 layers of SnSe is much thicker than the penetration depth, it is unlikely that the surface states of \ce{Bi2Se3} buffer layer will be observed. As shown in Fig.~\ref{fig:fig.S6}, at \textit{h$\nu$}=100eV, the cross section of Bi 5\textit{d} state is twice as big as that of the Sn 4\textit{d} state. Therefore, in Fig.2(a) of the main text, the peak at 28eV is assigned to chemically shifted Sn 4\textit{d} state with a chemical shift as Sn is the terminated surface.\\

\begin{figure}[h!]
\includegraphics[scale=0.45]{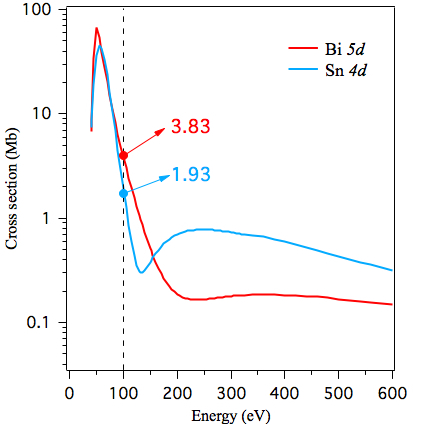}
\renewcommand{\thefigure}{S\arabic{figure}}
\caption{\label{fig:fig.S6}Atomic photoionization cross section for Bi 5\textit{d} and Sn 4\textit{d} subshells as a function of photon energy.} 
\end{figure}

\noindent \textbf{VII. DFT calculation of the localization length}\\

DFT calculations indicate that the decay lengths of the surface states at the $\Gamma$ point are short, \textit{i.e.,} $\sim$5 bilayers for a Se-terminated surface and $\sim$12 bilayers for a Sn-terminated surface. Thus 26 bilayers are sufficient to observe Dirac states at the $\Gamma$ point. However, our calculations suggest that this situation is different in the case of the Dirac states at the M point. The decay length of the surface states at M is much longer: it is about 20 bilayers for both types of surface terminations.

\begin{figure}[h!]
\includegraphics[scale=0.5]{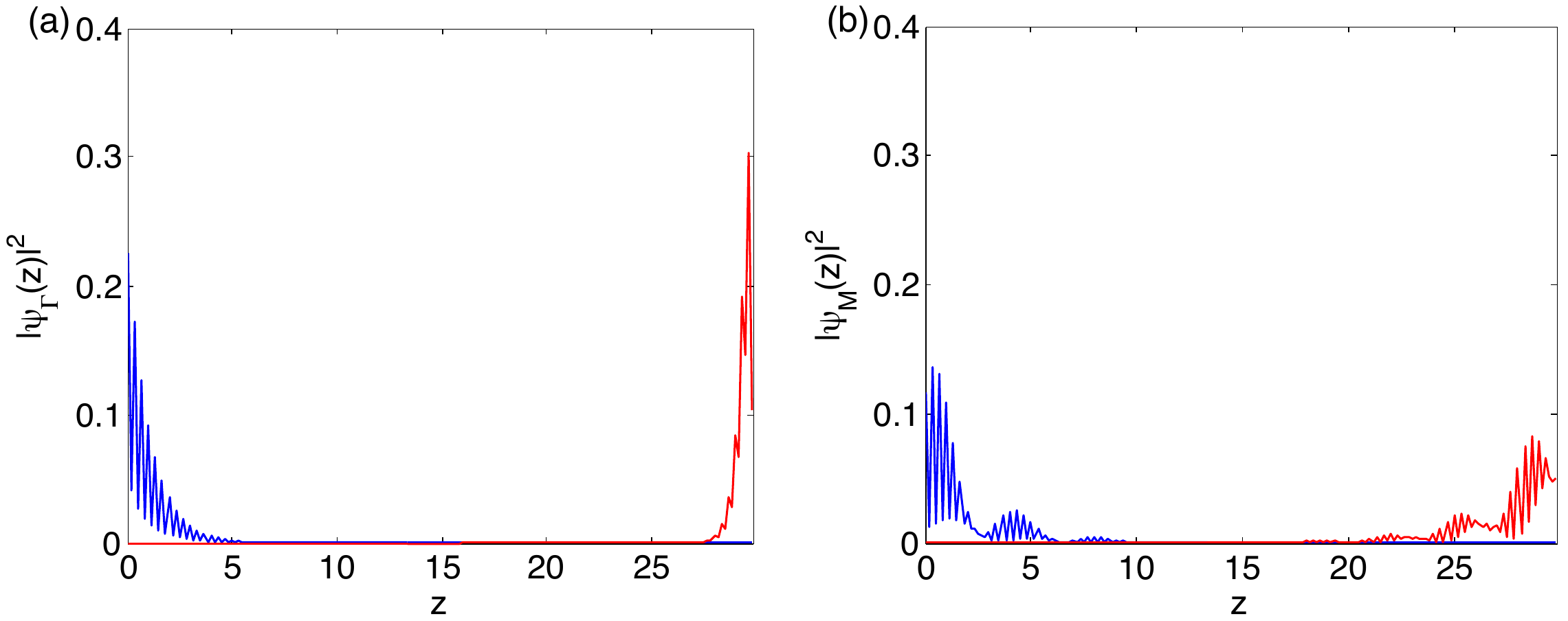}
\renewcommand{\thefigure}{S\arabic{figure}}
\caption{\label{fig:fig.S7}The localization length of surface states at (a) $\Gamma$ and (b) M as a function of the number of Sn-Se bilayers.} 
\end{figure}

\noindent \textbf{VIIi. Photon-energy-dependent ARPES measurement}\\

To verify the surface origin of the Dirac-like band, we carried out photon-energy-dependent ARPES measurement by tuning over a broad photon energy range of 20-75eV. As shown in Fig.~\ref{fig:fig.S7}, throughout the whole range, the spectra are nondispersive along $k_z$ direction, thus unambiguously confirming the presence of surface states.\\

\begin{figure}[h!]
\includegraphics[scale=0.45]{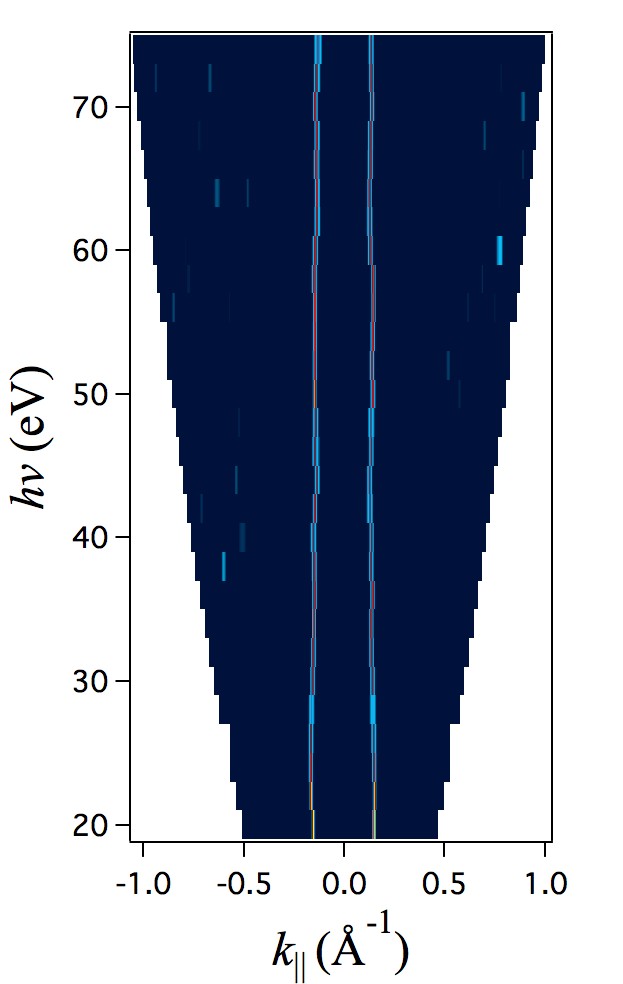}
\renewcommand{\thefigure}{S\arabic{figure}}
\caption{\label{fig:fig.S8}Second derivative plot of the photoemission spectral} 
{intensity as a function of momentum and photon energies.}
\end{figure}

\noindent \textbf{IX. DFT calculation using hybrid functional}\\

The bandgap size of bulk SnSe is further verified using DFT calculation with hybrid functionals (HSE). As shown in Fig.~\ref{fig:fig.S9}, the calculated bands using PBE functional (red) and HSE functional (blue) are very similar. The reason why DFT calculation underestimates the gap size of bulk SnSe is unclear at this time.\\

\begin{figure}[h!]
\includegraphics[scale=0.55]{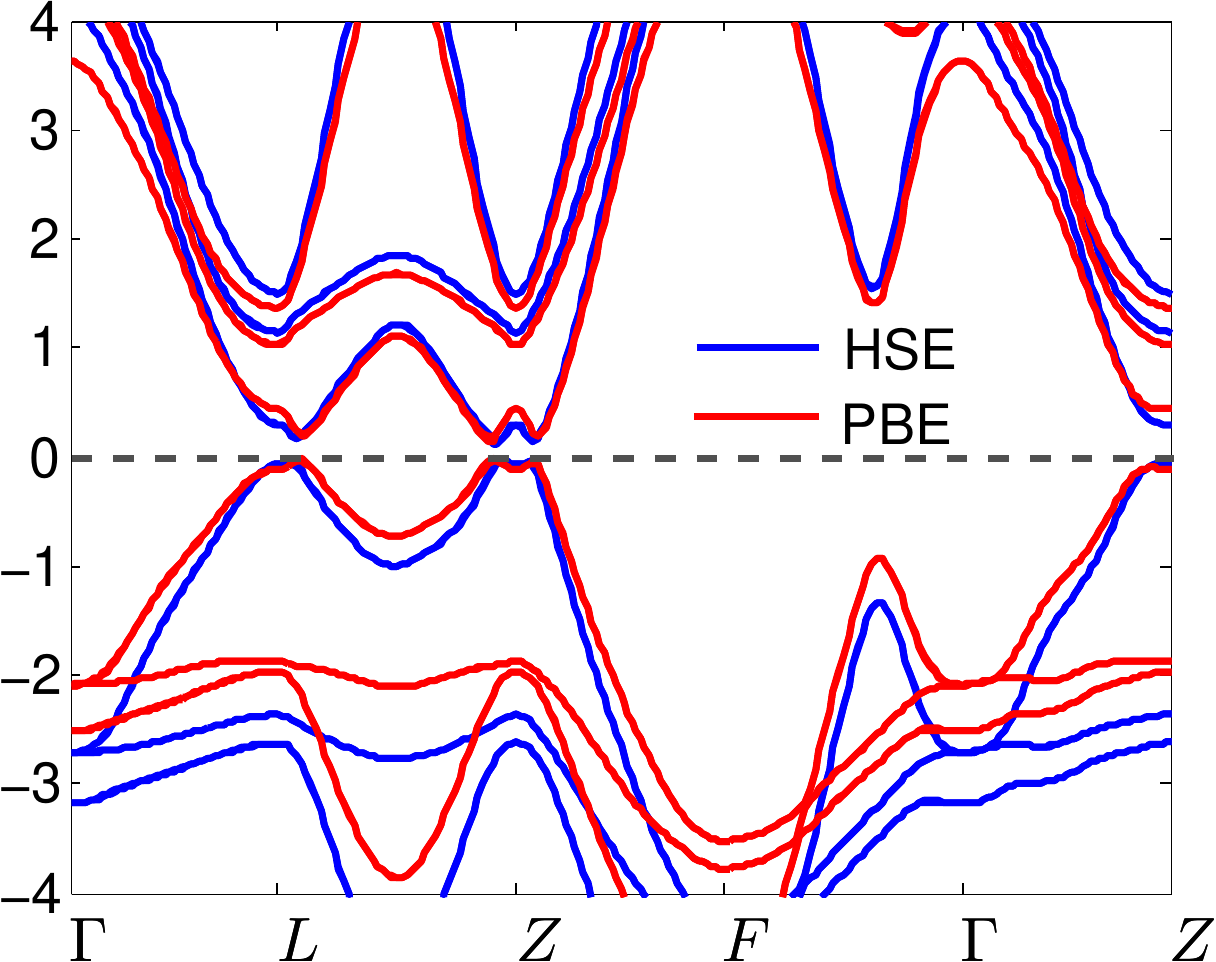}
\renewcommand{\thefigure}{S\arabic{figure}}
\caption{\label{fig:fig.S9} DFT calculated bands of bulk SnSe using PBE functional (red) and HSE functional (blue), respectively.} 
\end{figure}

\end{document}